\documentclass[pra,twocolumn,showpacs,superscriptaddress]{revtex4}
\usepackage{amsmath}
\usepackage{latexsym}
\usepackage{graphicx}
\usepackage{color}
\usepackage{bm}
\usepackage{float}
\usepackage{bbold}
\usepackage{amsfonts,amssymb}
\usepackage[normalem]{ ulem }
\usepackage{soul}
\usepackage{verbatim}
\usepackage{booktabs}

	\definecolor{blue-violet}{rgb}{0.54, 0.17, 0.89}
\definecolor{chromeyellow}{rgb}{1.0, 0.65, 0.0}
\definecolor{caribbeangreen}{rgb}{0.0, 0.8, 0.6}
\definecolor{light-blue}{rgb}{0.6,0.6,1}
\definecolor{cadmiumgreen}{rgb}{0.0, 0.42, 0.24}
\definecolor{Blue}{rgb}{0,0,1}
\definecolor{Red}{rgb}{1,0,0}

\newcommand{\N}[1]{{\mathcal N}_{#1}}
\DeclareMathOperator{\tr}{Tr}
\usepackage{hyperref} 
\usepackage[all]{hypcap}
\hypersetup{pdfauthor = {First Author},pdftitle = {Sending classical information via  noisy channels in superposition of causal orders}}

\begin{document}

\title{Sending classical information via  three noisy channels in superposition of causal orders}
\author{Lorenzo M. Procopio}

\thanks{Corresponding author: lorenzo.procopio@c2n.upsaclay.fr}
\affiliation{Centre for Nanoscience and Nanotechnology, C2N, CNRS, Universit\'e Paris-Sud, Universit\'e Paris-Saclay, 10 Boulevard Thomas Gobert, 91120 Palaiseau, France}

\author{Francisco Delgado}
\affiliation{Tecnologico de Monterrey, Escuela de Ingeniería y Ciencias, Carr. a Lago de Guadalupe km. 3.5, Atizap\'an, Estado de M\'exico, M\'exico, 52926}

\author{Marco Enr\'iquez}
\affiliation{Tecnologico de Monterrey, Escuela de Ingeniería y Ciencias, Carr. a Lago de Guadalupe km. 3.5, Atizap\'an, Estado de M\'exico, M\'exico, 52926}

\author{Nadia Belabas}
\affiliation{Centre for Nanoscience and Nanotechnology, C2N, CNRS, Universit\'e Paris-Sud, Universit\'e Paris-Saclay, 10 Boulevard Thomas Gobert, 91120 Palaiseau, France}

\author{Juan Ariel Levenson}
\affiliation{Centre for Nanoscience and Nanotechnology, C2N, CNRS, Universit\'e Paris-Sud, Universit\'e Paris-Saclay, 10 Boulevard Thomas Gobert, 91120 Palaiseau, France}

\date{\today}

\begin{abstract}
In this work, we study the transmission of classical information through three completely depolarizing channels in superposition of different causal orders. We thus introduce the quantum 3-switch as a resource for quantum communications. We perform a new kind of quantum control that was not accessible to the previously treated two-channel case. The fine and full quantum control achieved using selected combinations of causal orders let us uncover new features: non monotonous behavior on the transmission of information with respect to the number of causal orders involved, and different values of the  transmission of information depending on the specific combinations of  causal orders considered. Our results are a stepping stone to assess efficiency of coherent quantum control and optimize resources in the implementation of new indefinite causal structures. Finally, we suggest an optical implementation using standard telecom technology to test our predictions. 
\end{abstract}
\maketitle

\section{Introduction}

Contrary to  intuition, two fully noisy quantum channels  can still transmit classical information when they are  combined in a superposition of spatial or temporal trajectories, e.g., in a superposition of causal orders  where the order of application of the channels is indeterminate \cite{ebler2018enhanced}. This counter-intuitive effect is impossible to achieve using either one channel alone or a cascade of such fully noisy channels in a definite causal order. Causal activation of communication was invoked to explain this  result and also demonstrated to enable transmission of quantum information \cite{salek2018quantum}, even with a zero capacity channel \cite{chiribella2018indefinite}. Experimental realizations have recently tested this new activation of communication  \cite{goswami2018communicating,guo2018experimental}. In addition, the transmission of classical information has also been evidenced  controlling the path followed by the target system through one of two fully noisy channels with a quantum superposition  \cite{abbott2018communication}. The physical origin of each kind of communication enhancement through
quantum coherent control is thus currently the matter of stimulating discussions  \cite{chiribella2019quantum, guerin2019communication}.  
Interestingly, a generalization of quantum Shannon theory was proposed in \cite{chiribella2019quantum} encompassing both origins for enhancement. The well-established quantization of the internal degree of freedom of the information and/or channels is presented as a first level. The quantization of external degree of freedom, i.e. connections between channels, either through superposition of causal orders or superposition of paths, is considered as a second quantization level of the quantum Shannon theory of information. 

As an example of superposition of causal orders, the  operation known as  quantum switch has been initially designed by Chiribella et al \cite{Chiribella2013}. This primitive  has subsequently been theoretically proposed as a novel resource for applications to quantum information theory \cite{chiribella2012perfect,Araujo2014}, quantum communication complexity \cite{guerin2016exponential}, quantum communication \cite{salek2018quantum, procopio2019communication}, non-local games \cite{oreshkov2016causal} and quantum metrology \cite{zhao2019advantage,mukhopadhyay2018superposition}. Moreover, the quantum switch has been  implemented experimentally  \cite{procopio2015experimental,rubino2017experimental,goswami2018indefinite, guo2018experimental,wei2019experimental}. In the quantum switch, a target system $\rho$ undergoes a superposition of two different causal orders of application of two quantum channels. A control system $\rho_c$ is used to route target system; the state $\rho_c=\left| 1 \right>\left< 1 \right|$ encodes for  order where channel one is applied before channel two  while the state $\rho_c=\left| 2 \right>\left< 2 \right|$ encodes for  channel one  after channel two.  By placing $\rho_c$ in superposition, i.e.  $\rho_c=\left| + \right>\left< + \right|$, where $\left| + \right>_c= \frac{1}{\sqrt{2}}(\left| 1 \right> + \left| 2 \right> )$,   $\rho$ shall experience both orders simultaneously.

By increasing the number $N$ of quantum channels, the number of possible causal orders increases as $N!$. This, in turn, provides a large number of indeterminations, i.e. superposition of causal orders. The partial or total use of indeterminations in a quantum $N$-switch becomes a rich quantum resource, as we demonstrate in this paper for $N=3$. Indeed, this wealth of combinations of causal orders has been not yet exploited as it does not appear when only two channels are used in a quantum 2-switch.
In a recent work \cite{procopio2019communication}, we  presented a general procedure to study the transmission of classical information through $N$ quantum noisy channels using the quantum $N$-switch.  We use it  here the transmission of information beyond the two-channel paradigm resulting from fine quantum control of three noisy channels. We uncover new features of the transmission of information. Furthermore, the $N=3$ case  is  attainable with nowadays technologies and we sketch an optical implementation.

The paper is organized as follows. Section \ref{preliminaries} introduces the quantum 3-switch. It gives the necessary formalism and results to investigate the transmission of classical information by three fully  noisy quantum channels. We follow the formalism described in Ref. \cite{procopio2019communication} and obtain the quantum 3-switch matrix  to calculate the Holevo information, which quantifies transmission of information \cite{holevo1998capacity}. In Section \ref{transmission3}, we discuss the possible combinations of causal order that one can use to transmit information with three channels, and we analyze the effects on the transmission of information by superimposing $m$ causal orders, where $m=1,2,\dots,3!$. We further get and compare the Holevo information  as a function of the number of causal orders involved in the superposition. In Section \ref{implementation3}, we suggest an optical implementation for the quantum 2-switch and quantum 3-switch to test our predictions. Finally, in Section \ref{conclusion} we give  conclusions and perspectives of our work.
\begin{figure*}[ht]
	\begin{center} 
		\scalebox{.63}{\includegraphics{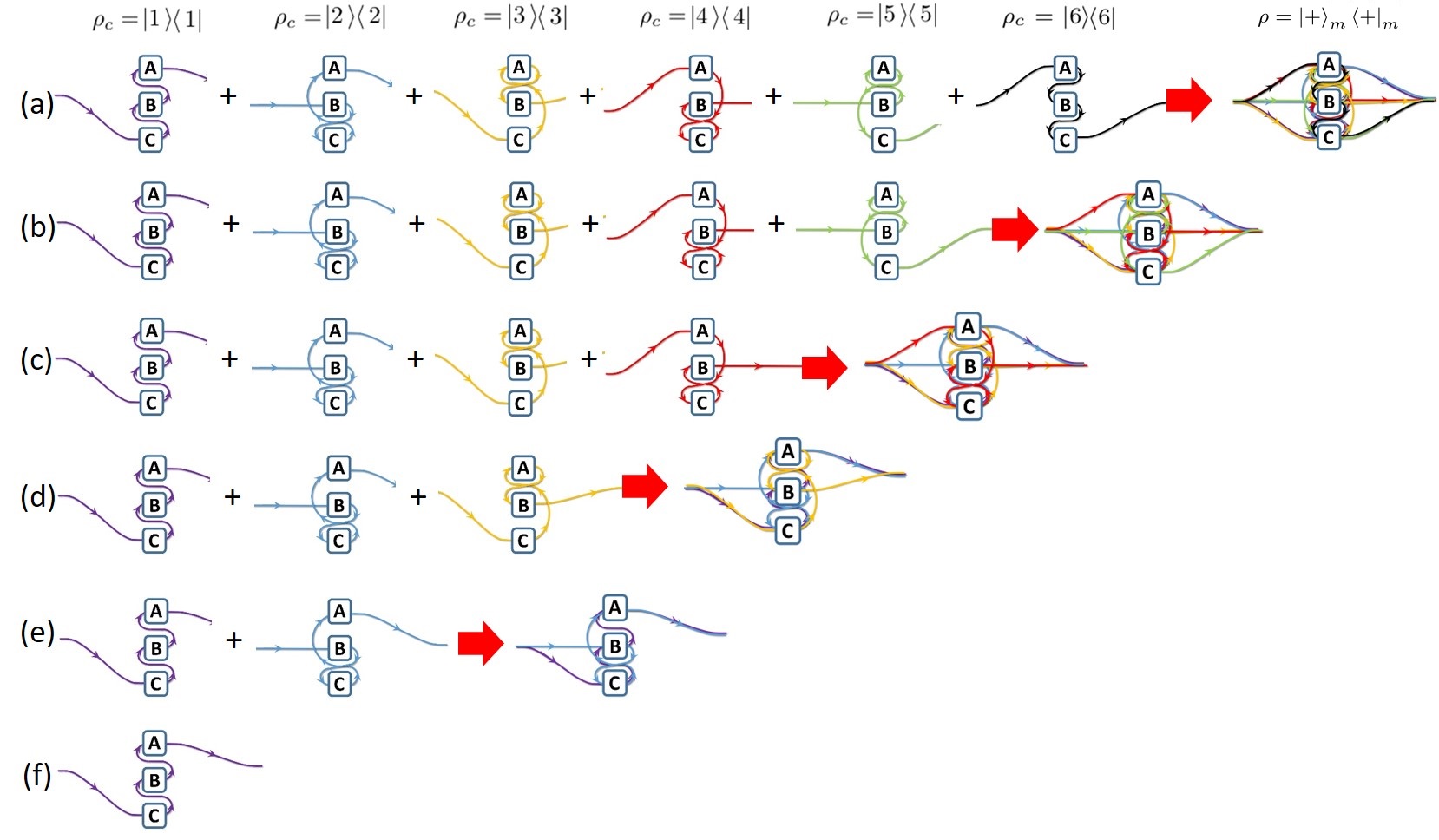}}
		\caption
		{\footnotesize \textbf{Tripartite quantum switch}. In the quantum switch with three parties, $\mathbf{A}$, $\mathbf{B}$ and $\mathbf{C}$, there are $\binom{3!}{m}$ (with $m=1,2,\ldots,6$ ) available superpositions of causal orders. Here we represent $\mathbf{A}$, $\mathbf{B}$ and $\mathbf{C}$ as three different quantum channels $\N{1}$, $\N{2}$ and $\N{3}$ respectively.  $\rho_c=\left| 1 \right>\left< 1 \right|$ encodes a causal order  $\N{1}\circ\N{2}\circ\N{3}$, i.e. $\N{3}$ is applied first to the target system $\rho$; $\rho_c = \left| 2 \right>\left< 2 \right|$ encodes $\N{1}\circ\N{3}\circ\N{2}$; $\rho_c = \left| 3 \right>\left< 3 \right|$ encodes $\N{2}\circ\N{1}\circ\N{3}$;  $\rho_c = \left| 4 \right>\left< 4 \right|$ encodes $\N{2}\circ\N{3}\circ\N{1}$;  $\rho_c = \left| 5 \right>\left< 5 \right|$ encodes $\N{3}\circ\N{1}\circ\N{2}$;   $\rho_c = \left| 6 \right>\left< 6 \right|$ encodes $\N{3}\circ\N{2}\circ\N{1}$; finally, if $\rho=\left| + \right>_{m}\left< + \right|_{m}$,  where $\left| + \right>_{m}= \frac{1}{\sqrt{m}}\sum_{k=1}^m \left| k \right>$  is  a superposition of  $m$ different causal orders. (a) For $m=6$ causal orders, there is only 1 combination to superimpose six definite causal orders.   (b) For $m=5$ causal orders, there are $\binom{3!}{5}=6$ combinations to superimpose five definite causal orders.  (c) For $m=4$ causal orders, there are $\binom{3!}{4}=15$ combinations to superimpose four definite causal orders. (d) For $m=3$ causal orders, there are $\binom{3!}{3}=20$ combinations to superimpose three definite causal orders. (e) For $m=2$ causal orders, there are $\binom{3!}{2}=15$ combinations to superimpose two definite causal orders.  (f) For $m=1$ causal order, depending on $\rho_c$, we have $3!$ possibilities to combine the channels in a definite causal order. Figures from (b) to (e)  illustrate a single partial superposition among all possible combinations of causal orders of Table~\ref{Table2}, see main text.}
		\label{Figura2}
	\end{center}
\end{figure*}
\section{Tripartite Quantum Switch}\label{preliminaries}

Three parties $\mathbf{A}$, $\mathbf{B}$ and $\mathbf{C}$ can be in $\sum_{m=2}^{N!}\binom{N!}{m}$ superpositions of $m$ causal orders with $N=3$, where $\binom{n}{r}= \frac{n!}{r!(n-r)!}$ is the binomial coefficient. Figure \ref{Figura2} describes possible superpositions of causal orders. In the context of quantum communications, $\mathbf{A}$, $\mathbf{B}$ and $\mathbf{C}$ are three  quantum noisy channels $\N{1}$, $\N{2}$ and $\N{3}$ respectively. We  model the action of a  noisy channel $\mathcal{N}$ on a qudit ($d$-dimensional) system $\rho$ as a depolarising quantum channel \cite{nielsen2002quantum}
\begin{equation} \label{depch2}
\mathcal{N}(\rho) = q \rho + (1-q) {\tr} [\rho] \frac{\mathbb{1}_t}{d}, 
\end{equation}

\noindent where $\mathbb{1}_t$ is the identity operator, which represents a maximally mixed state for the target system when the  channel is completely depolarizing., i.e., $q=0$. In this work, we will focus on the transmission of information through  completely depolarizing channels. We use the Kraus decomposition  $\mathcal{N}(\rho)=\sum_{i} K_{i}  \rho K_{i}^{\dagger}$ to mathematically represent the action of a channel ${\mathcal N}$ on the quantum state $\rho$ such that $\sum_{i} K_{i}  K_{i}^{\dagger}=\mathbb{1}_t$ \cite{nielsen2002quantum}.

For three noisy channels $\mathcal{N}_1$, $\mathcal{N}_2$ and $\mathcal{N}_3$, the control system $\rho_c=\left| \psi_c \right>  \left< \psi_c \right|  $  coherently controls the target system $\rho$ via $3!$ states, see Fig.~\ref{Figura2} top. The quantum state of the control system writes 
\begin{align} \label{Control3}
\left| \psi_c \right>
& = \sum_{n=1}^6\sqrt{ P_k} \left| k \right>
\end{align}

\noindent where $P_k$ is the probability associated to the definite causal order $\left| k \right>\left< k \right|$ such that $ \sum_{k=1}^{6} P_k=1$.  
If $\rho_c$ is at the state $\left| 1 \right>$, then the order to apply the channels will be $\N{1}\circ\N{2}\circ\N{3}$. Likewise, if $\rho_c$ is on the states $\left| 2 \right>$, $\left| 3 \right>$, $\left| 4 \right>$, $\left| 5 \right>$ or $\left| 6 \right>$,   the orders will be $\N{1}\circ\N{3}\circ\N{2}$,
  $\N{2}\circ\N{1}\circ\N{3}$,
  $\N{2}\circ\N{3}\circ\N{1}$,
 $\N{3}\circ\N{1}\circ\N{2}$ and $\N{3}\circ\N{2}\circ\N{1}$ respectively. Setting  the control state as in equation (\ref{Control3}), yields a superposition of several causal orders with their respective weights $P_k$. We refer to this type of superposition as the quantum 3-switch (Q3S) which is an extension of the quantum 2-switch (Q2S).

If the Kraus operators of the channels $\mathcal{N}_1$, $\mathcal{N}_2$ and $\mathcal{N}_3$ are $\{K_{i}^{(1)}\}$, $\{K_{j}^{(2)}\}$ and $\{K_{k}^{(3)}\}$ respectively, then the  Kraus operators $\mathcal{W}_{ijk}$ of the full quantum 3-switch channel is
\begin{align} \label{Kraus3}
\mathcal{W}_{ijk}
& = K_{i}^{(1)}K_{j}^{(2)}K_{k}^{(3)} \left| 1 \right>\left< 1 \right|+ K_{j}^{(1)}K_{k}^{(3)}K_{i}^{(2)} \left| 2 \right>\left< 2\right| \nonumber \\
& + K_{k}^{(2)}K_{i}^{(1)}K_{j}^{(3)} \left| 3 \right>\left< 3\right| +
K_{i}^{(2)}K_{k}^{(3)}K_{j}^{(1)} \left| 4 \right>\left< 4 \right| \nonumber\\
& + K_{j}^{(3)}K_{i}^{(1)}K_{k}^{(2)} \left| 5 \right>\left< 5\right| + K_{k}^{(3)}K_{j}^{(2)}K_{i}^{(1)} \left| 6 \right>\left< 6\right|.
\end{align}

The action of the quantum $3$-switch  ${\mathcal S}(\N{1},\N{2},\N{3})$ over an input $\rho \otimes \rho_c $  can be expressed through the  Kraus operators $\mathcal{W}_{ijk}$ as
\begin{equation}\label{Krausg}
\mathcal S(\N{1},\N{2},\N{3}) \left( \rho \otimes \rho_c \right)  =  \sum_{ijk} \mathcal{W}_{ijk}\left( \rho \otimes \rho_c \right.) \mathcal{W}_{ijk}^\dagger.
\end{equation}

Following the procedure described in Ref.  \cite{procopio2019communication}, we found  the quantum 3-switch matrix is a $6\times6$ block-symmetry matrix
\begin{equation}\label{Matrix3}
\begin{array}{lll}
\mathcal S=\left(\begin{array}{cccccc} 
\mathcal{A}_1 &\mathcal{B} & \mathcal{C}&\mathcal{D}&\mathcal{E}&\mathcal{F}\\
\mathcal{B} &\mathcal{A}_2 &\mathcal{G}&\mathcal{H}&\mathcal{I}&\mathcal{J}\\ 
\mathcal{C} &\mathcal{G} & \mathcal{A}_3&\mathcal{K}&\mathcal{L}&\mathcal{M}\\ 
\mathcal{D} &\mathcal{H} & \mathcal{K}&\mathcal{A}_4&\mathcal{N}&\mathcal{P}\\
\mathcal{E} &\mathcal{I}& \mathcal{L}&\mathcal{N}&\mathcal{A}_5&\mathcal{Q}\\
\mathcal{F}&\mathcal{J} & \mathcal{M}&\mathcal{P}&\mathcal{Q}&\mathcal{A}_6\end{array}\right),
\end{array}
\end{equation}
\noindent where $\mathcal{S}\equiv S(\N{1},\N{2},\N{3}) \left( \rho \otimes \rho_c \right)$, diagonal and off-diagonal elements are $d\times d$ matrices, linear combinations of the identity matrix $\mathbb{1}_t$ and the target system is the density matrix $\rho$, see  Appendix A of Ref. \cite{procopio2019communication}.
\noindent This quantum 3-switch matrix (\ref{Matrix3}) provides the output state of the quantum switch as a function of all the parameters involved: the dimension of the target system $d$, the depolarization strengths $q_i$'s, the probabilities $P_k$ and the density  matrix $\rho$.

We compute the Holevo information $\chi(\mathcal{S})$ of  the full quantum 3-switch channel $\mathcal{S}(\N{1},\,\N{2},\N{3})$  through a generalization of the mutual information,  see  Refs \cite{wilde2013quantum,ebler2018enhanced}.  $\chi(\mathcal{S})$ is found by maximizing the mutual information, and it can be shown that maximization over pure states $\rho$ for the target system is sufficient \cite{wilde2013quantum}. The Holevo information is then given by
\begin{equation}\label{Gholevo}
\chi \big({\mathcal S}\big) = \log d + H({\tilde \rho}_c^{(3)}) - H^\text{min}({\mathcal S})
\end{equation}
\noindent where $d$ is the dimension of the target system, $H^\text{min}(\mathcal{S})$ is the minimum of the entropy over pure states $\rho$ at the output of the channel $\mathcal{S}$. $H({\tilde \rho}_c^{(3)})$ is the Von-Neumann entropy of the output state of the control system $\tilde \rho_c^{(3)}$ for three channels which was found in Ref. \cite{procopio2019communication} Eq. (44).  The diagonalization and minimization of $H^\text{min}({\mathcal S})$ are performed numerically. 

\section{Fine quantum control of $N=3$ channels}\label{transmission3}
\subsection{Information transmission for full and partial superpositions}

\begin{figure}
	\begin{center} 
		\scalebox{.5}{\includegraphics[width = .9\textwidth]{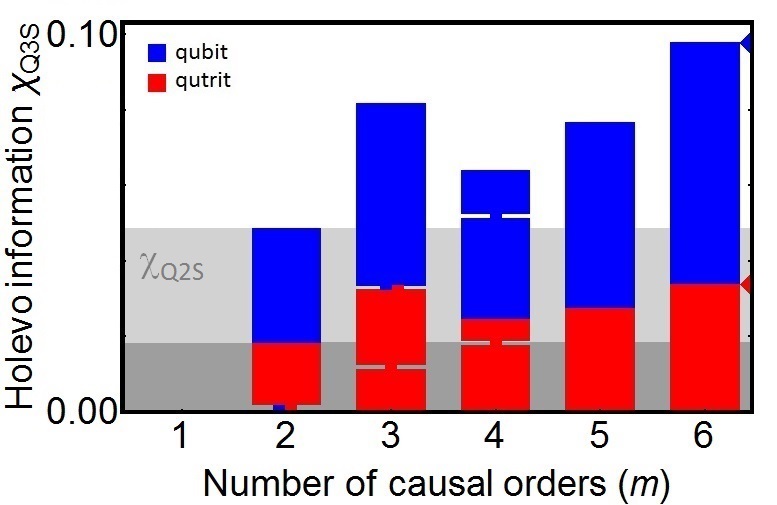}}
		\caption{\footnotesize \textbf{ Superposition of  $m$ causal orders}.  Holevo information $\chi_{\text{Q3S}}$ as the number of causal orders $m$ involved in $\rho_c$ is varied, for dimension $d=2$ (blue) and $d=3$ (red) of $\rho$.  The Holevo information can take 2 different values $\chi^\text{max}$ and $\chi^\text{min}$ when there is a superposition of  $m=$2, 3 and 4  causal orders, depending on the chosen causal orders in the superposition. $\chi^\text{max}$ is indicated by the top end of each color bar (for each $d=2, 3$) and $\chi^\text{min}$ is given by the position of the clear horizontal line bar with a colored centered dot. The numerical values of $\chi^\text{max}$ and $\chi^\text{min}$ are given in the Table \ref{table1}. For a fixed number of causal order $m$, the Holevo information  decreases as $d$ increases. Note that for superposition of two causal orders, $\chi^{\text max}$ is equal to the value of Holevo information of two channels $\chi_{\text{Q2S}}$. See main text for explanation. The two triangles correspond to the values obtained  in Ref. \cite{procopio2019communication}. } \label{superpositions}
	\end{center}
\end{figure}

As the number of channels increases, the number of possible causal orders increases as well : $2$ for $N=2$, $6$ for $N=3$, following a $N!$ law. This in turn increases the number of possible superpositions of combinations.  We analyze here in details the Holevo information with respect to these superpositions in the case of three channels.Note that this constitutes a new issue, since for $N=2$ there is no room to play with superposition of different causal orders.

Each definite causal order is associated to the control state $\left| k \right>\left< k \right|$ with probability $P_k$.  We  consider  all possible superposition of different causal orders  with equally weighted probabilities $P_k$ i.e. for each superposition of $m\in[\![1;N!=6]\!]$ causal orders, we fix  $m$ probabilities  $P_k$ to $\frac{1}{m}$ and the rest of $P_k$'s to zero.  We restrict our analysis to the case in which the three channels are completely depolarizing, i.e., $q_1=q_2=q_3=0$.  

There are $\binom{3!}{m}$ possible superpositions of  $m$ different causal orders for three channels. For $m=1$ causal order, there are six possible configurations, each in a specific definite causal order. We label each causal order  encoded by $\rho_c = \left| k \right>\left< k \right|$  with $P_k$  in the full list of combinations shown in Table~\ref{Table2}. We only display one definite causal order identified with $P_1$ and partial superpositions $P_1P_2$, $P_1P_2P_3$, $P_1P_2P_3P_4$ and $P_1P_2P_3P_4P_5$ Fig.~\ref{Figura2}(f)-(b) respectively. Fig.~\ref{Figura2}(a) displays the full superposition taking into account all orders.  In total we analyze $57=\sum_{m=2}^{6}\binom{3!}{m}$ superpositions of combinations of different causal orders for a fixed dimension $d$ of the target state.  Fig.~\ref{superpositions} gives the transmission of information  $\chi_{\text{Q3S}}$  for all possible superpositions for $d=2$ (blue) and $d=3$ (red). Table~\ref{Table2},  gives the possible combinations of $m$ causal orders related to the values of Fig.~\ref{superpositions}.

The main findings in Fig.~\ref{superpositions} are:

\begin{itemize}
	\item For a fixed dimension $d$, the transmission of information mostly increases as the number  $m$ of involved causal orders increases.
	\item This behavior is nonetheless not strictly monotonous. It is therefore unnecessary to waste resource to go from the $m=3$ to the $m=4$ case as the $3$-order-combination achieves better information transmission.
	\item The analysis of limited number of combinations is a novelty that could not be studied in the previous $N=2$ works. For $m=2$, $m=3$ and $m=4$, the possible combinations fall into 2 categories whether they transmit $\chi^\text{min}$ or $\chi^\text{max}$, $\chi^\text{min}<\chi^\text{max}$, see Table~\ref{table1}  that exhibits the values of Holevo information for $m=1,2,\ldots,6$ causal orders and distinct values of $d$, the	dimension of the target state.
	\end{itemize}

\begin{table}[ht]\label{tab2}
	\centering 
	\fontsize{8}{7.2}
	\begin{tabular}{|c|c c| c c|} 
		\hline
		\hline 
		$m$ &$\chi^\text{max}$ &  & $\chi^\text{min}$  & \\ [0.5ex]
		& $d=2$  & $d=3$ & $d=2$ & $d=3$\\ [0.5ex]
		\hline 
		1 & 0&0 & 0 &0 \\ 
		2 & 0.0487 & 0.0183 &0 &0 \\
		3 & 0.0817 & 0.0325 &0.0333&  0.0122\\
		4 & 0.0640 & 0.0246 & 0.0524 & 0.0186\\
		5 & 0.0766 & 0.0275 &-  &-\\
		6 & 0.0980&0.0339  & -&- \\ [1ex] 
		\hline
		\hline 
	\end{tabular}
	\caption{$\chi^\text{max}$ and $\chi^\text{min}$ values of the Holevo information of Figure \ref{superpositions} for all causal orders in $N=3$.}
	\label{table1} 
\end{table}

In more details and for increasing $m$,

\noindent {\it Case $m=1$.} In the case of $m=1$ the Holevo information  $\chi_{\text{Q3S}}$ reduces to that of a definite causal order scheme, i.e.  no information can be extracted. 

\noindent {\it Case $m=2$.} For $m=2$ the $15$ possible combinations are evaluated to be one of two values $\chi^\text{min}=0$ and $\chi^\text{max}$, the maximal one $\chi^\text{max}$ is endorsed by $6$ superpositions of different combinations (see Table~\ref{Table2}) and it coincides with the Holevo information obtained exploiting fully the two channel configuration i.e., $\chi^\text{max}=\chi_{\text Q2S}$ $(q_i=0)$\cite{ebler2018enhanced}. 
\noindent {\it Case $m=3$.} By increasing the causal order resource exploitation for the three channels case to $m>2$, the Holevo information is enhanced. Note that 
for $m=3$ ($20$ possible selections for the superpositions involved in the control states), superposition activates information transmission for all combinations of causal orders, and transmission is maximal for two specific combinations for which the transmitted information is 1.67 times larger than the transmitted information using two channels. It is interesting to notice that it is possible to surpass the bound in the transmission of information for the quantum 2-switch, combining three causal orders instead of involving all causal orders in the quantum 3-switch. From the experimental point of view, this can help reducing the complexity of implementations.

\noindent {\it Case $m=4$.} For $m=4$, the Holevo information is smaller than those combinations of $m=3$ where the transmitted information is maximum.

\noindent {\it Case $m=5$.} The two values collapse into a single one for the 6 possible combinations associated to $m=5$. 

\noindent {\it Case $m=6$.} Remarkably, when the $3$-channel resources are fully exploited, for the single equally weighted combination of the $m=6$ case, the Holevo information is approximately two  times  that of the two channel configuration up to $d=10$ as it was found in Ref. \cite{procopio2019communication}.  

\subsection{Realistic quantum switch and efficiency of transmission of local and global combinations}

The case $m=2$ can be understood as follows. For those combinations of causal orders where superposition activation is on, i.e. $\chi_{\text Q3S}=\chi^\text{max}$, the quantum 3-switch is switching globally all channels, i.e. all channels are combined in such a way that they all have  changed positions in the ordering. In the particular case when there is information transmission $\chi_{\text Q3S}=\chi^\text{max}\neq0$, the quantum 3-switch  is in fact switching only two channels:  one channel $\mathcal{N}_i$ and another composite channel $\mathcal{N}_{jk}=\mathcal{N}_j\circ\mathcal{N}_k$. The quantum 3-switch thus indeed behaves as an effective  quantum 2-switch. Hence $\chi_{\text Q3S}(m=2)$ is equal to $\chi_{\text Q2S}.$

In contrast,  for those combinations where $\chi_{\text Q3S}=\chi^\text{min}=0$, the quantum 3-switch is switching locally only two individual channels  $\mathcal{N}_j$ and  $\mathcal{N}_k$ instead of globally switching all channels.  Indeed the $\chi_{\text Q3S}=\chi^\text{min}=0$ case is interesting to shed light on the seminal quantum 2-switch scheme. In fact, Q2S incorporates an implicit  channel linking the output of $\mathcal{N}_{2}$ to the input of $\mathcal{N}_{1}$. This identity channel has no loss in the $N=2$ picture as it would be otherwise a third channel. The transmission of information reported in Ref. \cite{ebler2018enhanced} for the Q2S could be attributed to this implicit channel. Our results for $N=3$ give the proper framework to make this third channel explicit. If this identity channel is a fully depolarizing  channel,  our result for $N=3$ and the corresponding $m=2$ case  shows that  indeed  no information  is transmitted.

The behavior of Fig.~\ref{superpositions} can  be summarized by noticing that the more the quantum switch switches channels, the more information is transmitted. It seems that the $m$ dependence of the Holevo information can  be tracked back comparing the number  of the combinations in the control state and their nature that we label local and global. As detail below Table~\ref{localglobal} illustrates our global and local switching denomination. It  gives the explanation for the Holevo values obtained from selected combinations switching three channels with $m=3$ and $m=4$ orderings. We shall present a more formally proof using  properties of the quantum switches matrix~(\ref{Matrix3}) in a future work.   

For $m=3$, the combination $P_3 P_1 P_2$  yields a low value for the Holevo information $\chi^\text{min}(m=3)$. In 2 subsets of 2 causal orders we only locally switch the channels.  We highlight this by putting colors on the fixed points which indicate this local switching. This $P_3P_1P_2$ combination only has 1 subset  of 2 causal orders $\{P_3,P_2\}$  that globally switch the channels without fixed point. In contrast, in the $P_1 P_4 P_5$ superposition there are 3 subsets of 2 causal orders to globally exchange all channels thus yielding $\chi^\text{max} (m=3)$. The global switching has no fixed points.

For $m=4$, the combination yielding $\chi^\text{min} (m=4)$ has $4$ highlighted fixed points, i.e., 4 pairs of local switching, or 2 pairs that globally switch the channels among the possible $6$ pairs. While there are only $3$ fixed points and $3$ possible combinations to globally switch the $3$ channels for the $P_1 P_5 P_2 P_4$ superposition yielding $\chi^\text{max} (m=4)$. Note that the number of combinations to globally switch the channels yielding the high values $\chi^\text{max} (m)$ of $m=3$ and $m=4$ are equal, however they amount to all possibilities for $m=3$, where some combinations only achieve local switching for $m$=4 which results in our interpretation in a decrease of transmitted information in the results shown in Fig.~\ref{superpositions}.

Our reasoning is independent of the dimension of the target state $d$. Note that indeed the Holevo information decreases as $d$ increases but the overall $m$ dependence is the same.

\begin{table}[ht]\label{localglobal}
	\centering 
	\fontsize{8}{7.2}
	\begin{tabular}{|cccc|cccc|} 
		\hline
		\hline 
		$\chi^{{\rm min}}$&$(m=3)$ & & & $\chi^{{\rm max}}$&$(m=3)$&& \\ [0.5ex]
		\hline 
		$\mathbf{P_3}$ &\color{blue}{$N_3$}&$N_1$&$N_2$&$\mathbf{P_1}$&$N_3$&$N_2$&$N_1$ \\ 
		$\mathbf{P_1}$ &\color{blue}{$N_3$}&$N_2$&\color{caribbeangreen}{$N_1$}&$\mathbf{P_4}$&$N_1$&$N_3$&$N_2$ \\
		$\mathbf{P_2}$ &$N_2$&$N_3$&\color{caribbeangreen}{$N_1$}&$\mathbf{P_5}$&$N_2$&$N_1$&$N_3$\\ [1ex] 
		\hline
		\hline
	\end{tabular}
	
	\vspace{2ex}
	
	\begin{tabular}{|cccc|cccc|} 
		\hline
		\hline 
		$\chi^{{\rm min}}$& $(m=4)$ & & & $\chi^{{\rm max}}$&$(m=4)$&& \\ [0.5ex]
		\hline 
		$\mathbf{P_3}$ &\color{blue}{$N_3$}&$N_1$&\color{caribbeangreen}{$N_2$}&$\mathbf{P_1}$&$N_3$&$N_2$&\color{caribbeangreen}{$N_1$} \\ 
		$\mathbf{P_1}$ &\color{blue}{$N_3$}&$N_2$&\color{blue-violet}{$N_1$}&$\mathbf{P_5}$&\color{chromeyellow}{$N_2$}&$N_1$&$N_3$ \\
		$\mathbf{P_2}$ &$N_2$&\color{chromeyellow}{$N_3$}&\color{blue-violet}{$N_1$}&$\mathbf{P_2}$&\color{chromeyellow}{$N_2$}&\color{blue}{$N_3$}&\color{caribbeangreen}{$N_1$}\\ 
		$\mathbf{P_4}$ &$N_1$&\color{chromeyellow}{$N_3$}&\color{caribbeangreen}{$N_2$}&$\mathbf{P_4}$&$N_1$&\color{blue}{$N_3$}&$N_2$\\[1ex] 
		\hline
		\hline
	\end{tabular}
	\caption{\textbf{Example of evaluation of combinations switching three channels}. We detail here the four examples of superposition of $m$ orders underlined in Table~\ref{Table2} and relate them to their high $\chi^\text{max} (m)$ or low $\chi^\text{min} (m)$ transfer of information, by evaluating the ratio of globally switching pairs among possible pairs in the superposition of $m$ orders. Colors indicate fixed points, see main text.}
	\label{localglobal} 
\end{table}

\section{Implementation}\label{implementation3}

\begin{figure*}[ht]
	\begin{center} 
		\scalebox{.7}{\includegraphics[width = 1\textwidth]{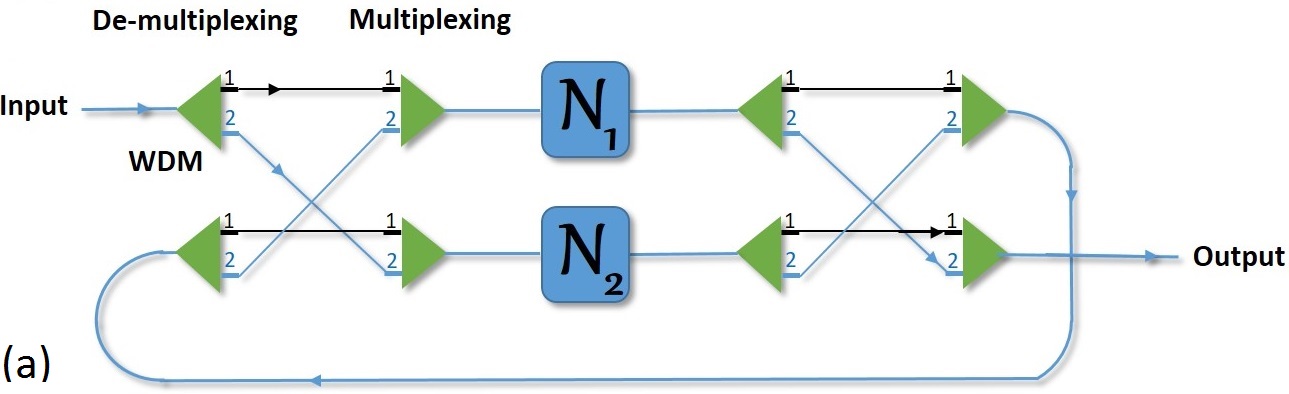}}\\
		\scalebox{.7}{\includegraphics[width = 1\textwidth]{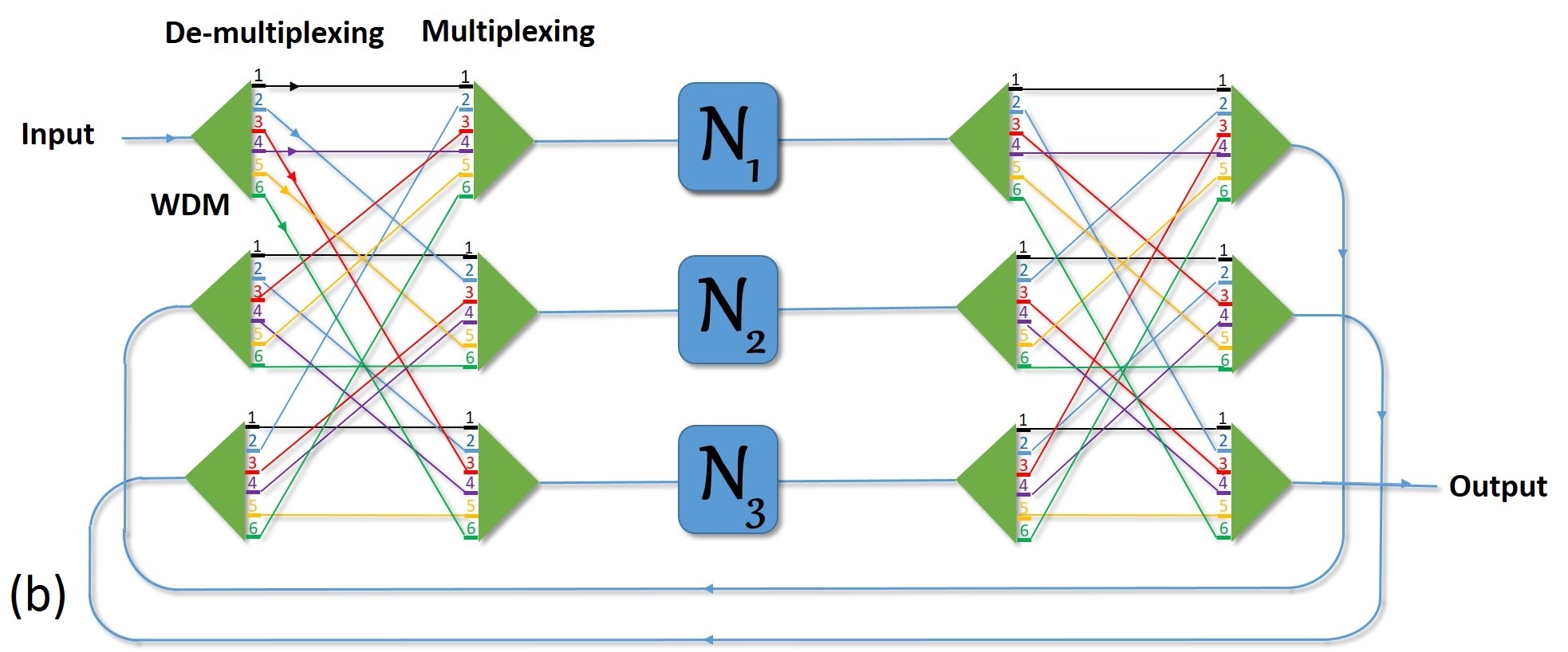}}\\
		\caption{\footnotesize \textbf{Optical proposal for the quantum $N$-switch channel.}  In both proposals,  a frequency-delocalized single photon with $N!$ frequencies is injected as an input to a series of wavelength division multiplexers and de-multiplexers (WDMs). Each  frequency is used to route the order of operation of the channels. At the end of the quantum switch, the frequencies are coherently multiplexed into the output mode. All color lines represents optical links which connect the WDMs and the channels $\mathcal{N}_j$. (a) For the quantum 2-switch, if the frequency is on mode 1 (black), the order to apply the channels  will be $\mathcal{N}_2\circ\mathcal{N}_1$. On the other hand, if the frequency is on mode 2 (blue), the order will be $\mathcal{N}_1\circ\mathcal{N}_2$. (b) For the quantum 3-switch, if the frequency is on mode 1 (black), the order to apply the channels  will be $\mathcal{N}_3\circ\mathcal{N}_2\circ\mathcal{N}_1$. If the frequency is on mode 2 (blue), the order will be $\mathcal{N}_1\circ\mathcal{N}_3\circ\mathcal{N}_2$. If the frequency is on mode 3 (red), the order will be $\mathcal{N}_2\circ\mathcal{N}_1\circ\mathcal{N}_3$. If the frequency is on mode 4 (purple), the order will be $\mathcal{N}_2\circ\mathcal{N}_3\circ\mathcal{N}_1$. If the frequency is on mode 5 (yellow), the order will be $\mathcal{N}_3\circ\mathcal{N}_1\circ\mathcal{N}_2$. Finally, if the frequency is on mode 6 (green), the order will be $\mathcal{N}_1\circ\mathcal{N}_2\circ\mathcal{N}_3$. By sending, in both cases (a) and (b), a single photon in a superposition of frequencies will have all causal orders simultaneously. Note that our optical proposal can be seen as the implementation for the  architecture proposed in \cite{Araujo2014}. We propose frequency encoding using off-the-shelf and mature telecom components.
		} \label{schemes}
	\end{center}
\end{figure*}

Fig.~\ref{schemes}  sketches our optical proposal to implement the quantum switch for two (a) and three (b) channels respectively. To implement  experimentally the quantum switch channel, two main ingredients are required: a control and target system.   Implementation of a quantum switch for $N\geq2$ thus faces several challenges: (i) Operations on the chosen quantum system should be applied only on the target system (dimension $d$), without disturbing the control system
(ii) The dimension $d_c$ of the control system, an appropriate quantum system to coherently control the order of operations, must be adapted to route all orders of the $N$ operations and grows as the number of permutations $N!$  (iii) As the number of operations $N$ grows, the experiment requires coherent control in a robust and scalable manner of dimension $d_c \times d$  which a priori grows as $N!\times N!$. In the existing experiments \cite{procopio2015experimental,goswami2018indefinite,wei2019experimental, guo2018experimental}, the control system has been realized using either the path or  polarization degrees of freedom of a single photon, and for the target system, it has been implemented using either the polarization or the transverse spatial mode of the same photon.  In all these implementations, fibered or in free space, the quantum switch is limited  due to the encoding  of  the  control system in a two dimensional space  and thus does not  scale up to more than $N=2$ quantum channels, although in \cite{wei2019experimental} the arrival time encodes a $d$-dimensional target system. 

We suggest to scale up the quantum $N$-switch to $N\geq2$ with the generation and manipulation of single photons at telecom-wavelength, in a frequency-comb structure \cite{kues2017chip,wang2019monolithic,mazeas2016high}. We propose  to use the frequency bin of a single photon from a comb as the control system for routing the causal order of the operation of the channels, i.e. each frequency bin of the single photon supports a different quantum state of $\rho_c$ and thus a different ordering of channels.  For the target system, we propose to use the time-bin \cite{wei2019experimental}  degree of freedom of the same photon going through the superposition of the channels.

In our suggested implementation, see Fig.\ref{schemes}, the input mode, a frequency-delocalized single photon with $N!$ frequencies is injected. The photon is firstly de-multiplexed using  wavelength division multiplexers (WDM), and then depending on the frequency, it is guided by selective optical links through the corresponding causal order $k$, which is acting separately on the time-bin degree of freedom.  At the end of the quantum switch, the frequencies of the single photon are coherently multiplexed into the output mode. Our scheme is feasible with telecom standard technology or in an integrated Silicon platform. It is only bounded by optical losses and has no fundamental limitation on implementing any arbitrary number $N$ of causal orders or increasing the dimensionality of the quantum systems involved. In practice this schemes requires robust and reliable filtering and perfect matching of fibered or integrated multiplexing and de-multiplexing to the frequency combs.\\

\section{Conclusion.}\label{conclusion}

 We have investigated quantum control of three fully-noisy channels in the context of quantum Shannon theory with superpositions of trajectories  in the specific case of  superposition of causal orders \cite{chiribella2019quantum,procopio2019communication}.
 
   We exhibit two different behaviors of the information transmission in the arbitrary $N$ case for two specific combinations of two causal orders. Our work on $N=3$ channels is a significant advance in quantum control of causal order : In contrast with the previous $N=2$ studies where only one combination of orders is accessible for quantum control, getting to $N=3$ provides 57 combinations. Beyond the mere increase of combinations, this opens-up a full quantum control through the game of local or global switches, something that is not possible for $N=2.$ We assessed them and exhibited the influence of the number and nature of the $m$ causal orders involved in those combinations on the Holevo information. We thus uncover new quantum features of indefinite causal structures with combinations that are more efficient than others.  Finally, we propose an implementation using standard telecom technology to test our findings experimentally. Our work is thus to our knowledge the first quantitative study of indefinite causal structures providing predictions in a multipartite scenario within a new paradigm  for the  quantum  information and quantum communications fields.



\emph{Acknowledgment.} L.M. Procopio wishes to thank  Fabio Costa for stimulating discussions on the optical implementation and  to Alastair Abbott for valuable
comments on a preliminary version of this manuscript.  F. Delgado and M. Enr\'iquez acknowledges the support from CONACYT and Escuela de Ingenier\'\i a y Ciencias of Tecnol\'ogico de Monterrey in the developing of this research work. L.M. Procopio acknowledges the support from the European Union's Horizon 2020 research and innovation
programme under the  Marie Sk\l{}ukodowska-Curie grant agreement No 800306.  This work is also supported by a public grant overseen by the French National Research Agency (ANR) as part of the ``Investissements d'Avenir'' program (Labex NanoSaclay, reference: ANR-10-LABX-0035) and by the Sitqom ANR project  (reference : ANR-SITQOM-15-CE24-0005).


\cleardoublepage

\def\thesection{}

\section*{Appendices}
\subsection{Combinations of superimposing $m$ causal orders}\label{A}

\begin{table}[ht]
	\begin{center}
		\fontsize{6}{7.2}
		\begin{tabular}{|c|c|c|c|c|c|c|c|}
			\hline
			$m=1$& $m=2$ & $m=3$& $m=4$& $m=5$&$m=6$  \\
			\hline
			\hline
			$P_1$ &$P_1$$P_2$&$\underline{P_1P_2P_3}$&\underline{$P_1P_2P_3P_4$}&$P_1P_2P_3P_4P_5$&$P_1P_2P_3P_4P_5P_6$\\
			$P_2$  &$P_1$$P_3$&$P_1P_2P_4$&$P_1P_2P_3P_5$&$P_1P_2P_3P_4P_6$&\\
			$P_3$ &\color{caribbeangreen}{$P_1$$P_4$}&$P_1P_2P_5$&\color{caribbeangreen}{$P_1P_2P_3P_6$}&$P_1P_2P_3P_5P_6$&	\\
			$P_4$ &\color{caribbeangreen}{$P_1$$P_5$}&$P_1P_2P_6$&\underline{\color{caribbeangreen}{$P_1P_2P_4P_5$}}&$P_1P_2P_4P_5P_6$&\\
			$P_5$ &$P_1$$P_6$&$P_1P_3P_4$&$P_1P_2P_4P_6$&$P_1P_3P_4P_5P_6$&\\
			$P_6$&\color{caribbeangreen}{$P_2$$P_3$}&$P_1P_3P_5$&$P_1P_2P_5P_6$&$P_2P_3P_4P_5P_6$&\\
			&$P_2$$P_4$&$P_1P_3P_6$&\color{caribbeangreen}{$P_1P_3P_4P_5$}&&\\
			&$P_2$$P_5$&\underline{\color{caribbeangreen}{$P_1P_4P_5$}}&$P_1P_3P_4P_6$&&\\
			&\color{caribbeangreen}{$P_2$$P_6$}&$P_1P_4P_6$&$P_1P_3P_5P_6$&&\\
			&$P_3$$P_4$&$P_1P_5P_6$&\color{caribbeangreen}{$P_1P_4P_5P_6$}&&\\
			&$P_3$$P_5$&$P_2P_3P_4$&$P_2P_3P_4P_5$&&\\
			&\color{caribbeangreen}{$P_3$$P_6$}&$P_2P_3P_5$&\color{caribbeangreen}{$P_2P_3P_4P_6$}&&\\
			&\color{caribbeangreen}{$P_4$$P_5$}&\color{caribbeangreen}{$P_2P_3P_6$}&\color{caribbeangreen}{$P_2P_3P_5P_6$}&&\\
			&$P_4$$P_6$&$P_2P_4P_5$&$P_2P_4P_5P_6$&&\\
			&$P_5$$P_6$&$P_2P_4P_6$&$P_3P_4P_5P_6$&&\\
			&&$P_2P_5P_6$&&&\\
			&&$P_3P_4P_5$&&&\\
			&&$P_3P_4P_6$&&&\\
			&&$P_3P_5P_6$&&&\\
			&&$P_4P_5P_6$&&&\\
			\hline
			\hline
		\end{tabular}
	\end{center}
	\caption{ \textbf{Table of possible combinations of $m$ causal orders.} $\rho_c=\sum_{k,k'=1}^{N!} \sqrt{P_k P_{k'}} \left| k \right>\left< k' \right|$ involves a superposition of $m$ causal orders, with $m$ different values. $P_i$ are non-zero probabilities.  There are $\binom{3!}{m}$ possibles configurations $m$ causal orders. The green color indicates which combinations  yield the maximum values $\chi^\text{max} (m)$ in   Fig.~\ref{superpositions} for $d=2$ and $d=3$. For simplicity we set for our estimates the non zero $P_i$ to be $\frac{1}{m}$. The underlined terms  correspond to the cases studied in Table~\ref{localglobal}. In Figures~\ref{Figura2}(b) to (e) we show only one combination of $P_k$'s from this table corresponding to a specific  partial superposition.}
	\label{Table2}
\end{table}

\cleardoublepage
\end{document}